\documentclass[reprint, amsmath, amssymb, aps]{revtex4-2}
\usepackage{graphicx}
\usepackage{dcolumn}
\usepackage{bm}
\usepackage{hyperref}
\usepackage{multirow}
\usepackage{amsmath}
\hypersetup{hypertex=true,
            colorlinks=true,
            linkcolor=blue,
            anchorcolor=blue,
            citecolor=blue}

\begin{document}
\preprint{APS/123-QED}

\title{Bridging Theory and Data: Correcting Nuclear Mass Models with Interpretable Machine Learning}

\author{Yanhua Lu, Tianshuai Shang, Pengxiang Du, Jian Li}
\email{E-mail:jianli@jlu.edu.cn}
 \affiliation{College of Physics, Jilin University, Changchun 130012, China.}
 \author{Haozhao Liang}
\affiliation{Department of Physics, Graduate School of Science,	The University of Tokyo, Tokyo 113-0033, Japan}
\affiliation{RIKEN Interdisciplinary Theoretical and Mathematical Sciences Program, Wako 351-0198, Japan}

\date{\today}

\begin{abstract}

Nuclear mass prediction is one of the core issues in nuclear physics research, yet it faces the challenge of small-sample datasets with high complexity. This study introduces the Kolmogorov-Arnold Network (KAN) into the refinement of nuclear mass models, proposing an efficient and interpretable solution. By constructing the KAN-WS4 hybrid model, the prediction accuracy is significantly improved (the root mean square error is reduced from 0.3 MeV to 0.16 MeV). Furthermore, leveraging the intrinsic interpretability of KAN, feature importance analysis reveals that the proton number is the most critical factor influencing residuals, indicating potential systematic biases in proton-related terms within existing theoretical models. The method's generality is demonstrated across five mass models. This study shows that KAN provides a novel approach to small-sample, high-complexity scientific problems. Its interpretability facilitates the data-driven discovery of physical laws, promising broad applicability to key nuclear physics issues.

\end{abstract}

\maketitle

\section{\label{sec1}introduction}

    Nuclear mass, one of the most fundamental properties of the nucleus, offers rich information on structural features such as deformation~\cite{PhysRevLett.96.042504, PhysRevC.96.014310} and shell effects~\cite{ramirez2012direct, wienholtz2013masses}. It is also widely used to extract effective nuclear interactions, including mean-field effective interactions and pairing interactions~\cite{PhysRevC.71.024312, PhysRevC.82.054319, PhysRevLett.81.3599}. Additionally, nuclear mass serves as a key input in nuclear physics for understanding stellar energy generation and the origin of elements in the universe~\cite{RevModPhys.29.547, MUMPOWER201686}. Due to the importance of nuclear masses, investigating them from both experimental and theoretical perspectives has long remained a crucial frontier research topic in nuclear physics.

    Considerable progress has been achieved in experimental measurements~\cite{Rau2020Penning, YAMAGUCHI2021103882}. The Atomic Mass Evaluation 2020 (AME2020) provides a comprehensive overview of nuclear mass measurements, evaluating and recommending the nuclear masses of over 3,000 nuclides~\cite{Huang_2021, Wang_2021}. On the theoretical side, the prediction of nuclear masses has a long history. In 1935, Weizs$\ddot{\rm{a}}$cker~\cite{weizsacker1935theorie, bethe1936nuclear} proposed the first semi-empirical formula for calculating nuclear masses, known as the Bethe–Weizs$\ddot{\rm{a}}$cker (BW) mass formula, which achieved a predictive accuracy of around 3 MeV. Semi-empirical formulas belong to the category of macroscopic models and provide a good description of masses for most nuclei. By incorporating microscopic shell corrections into such models, researchers developed macro–micro models such as the finite-range droplet model (FRDM)~\cite{PhysRevLett.108.052501} and the Weizsäcker–Skyrme (WS) model~\cite{WANG2014215}, which significantly improved the accuracy to approximately 0.3 MeV. Microscopic mass models based on nuclear density functional theory have also been developed, achieving an accuracy of approximately 0.5 MeV~\cite{VRETENAR2005101, MENG2006470, PhysRevLett.101.122502, PhysRevC.87.051303, PhysRevC.95.044301, XIA20181, 10.1143/PTP.113.785, pena2016relativistic, PhysRevLett.102.242501, PhysRevC.93.034337, ZHANG2022101488, GUO2024101661}. Although their accuracy in reproducing experimentally known masses is currently lower than that of macro–micro models, they are generally considered to provide more reliable extrapolations~\cite{PhysRevC.86.064324}. The theoretical models mentioned above are typically classified as global models. In contrast, models based on local systematics that relate properties of neighboring nuclides are termed as local mass models~\cite{PhysRevLett.16.197, PhysRevC.82.054317, PhysRevC.84.034311, PhysRevC.85.054303}. Such local models can achieve high predictive accuracy for nuclear masses, especially in regions where experimental data already exist.

    Despite considerable improvements in the predictive accuracy of nuclear mass models, current theoretical approaches still fall short of the requirements for studying exotic nuclear structures and astrophysical nucleosynthesis. For instance, a key phenomenon in nuclear astrophysics—the rapid neutron-capture process (r-process)—demands mass predictions with an accuracy on the order of 0.1 MeV~\cite{MUMPOWER201686}. Achieving such precision is therefore a key objective. Furthermore, high-precision mass data could even be leveraged to probe weak interactions, test CPT symmetry, and contribute to determining fundamental physical constants~\cite{RevModPhys.75.1021, BLAUM20061, atoms7010037}. Consequently, substantial potential remains for further refinement of existing nuclear mass models.

    In recent years, machine learning (ML) has emerged as a prominent approach for addressing complex problems, prized for its powerful learning capabilities. In the field of nuclear physics, ML is playing an increasingly important role with a wide range of applications~\cite{RevModPhys.94.031003, zongshuhewanbing2023}. These include predicting properties of nuclear ground and excited states, such as charge radii~\cite{Akkoyun_2013, Utama_2016, PhysRevC.101.014304, PhysRevC.102.054323, DONG2023137726, Lul3yu}, excited states~\cite{PhysRevLett.124.162502, BAI2021136147, WANG2022137154}, $\alpha$-decay~\cite{Rodríguez_2019, PhysRevC.108.014326}, $\beta$-decay~\cite{PhysRevC.99.064307, Li_2024}, charge density~\cite{shang2022prediction, PhysRevC.108.034315, PhysRevC.110.014308}, density functionals~\cite{YANG2023137870}, nuclear level density~\cite{PhysRevC.109.044325} and ground-state magnetic moments~\cite{Yuan_2021}. Notably, ML has been extensively employed for nuclear mass predictions~\cite{PhysRevC.93.014311, PhysRevC.96.044308, NIU201848, PhysRevLett.122.062502, PhysRevC.101.044307, PhysRevC.100.054311, PhysRevC.101.035804, PhysRevC.104.014315, PhysRevC.106.014305, PhysRevC.106.L021301, WU2021136387, PhysRevC.106.L021303, PhysRevC.109.034318, PhysRevC.111.014325, PhysRevC.111.024316}, contributing to improved accuracy in the description of nuclear masses.

    Most ML models, especially deep learning architectures, typically involve complex network with a large number of parameters. Training such models to high performance generally requires extensive datasets. For instance, image classification datasets like MNIST contain around 60,000 samples~\cite{6296535}, while ImageNet includes over 14 million samples~\cite{5206848}. In natural language processing, large-scale corpora comprise billions of tokens~\cite{devlin-etal-2019-bert}. When applied to scientific research, ML typically also requires hundreds of thousands of samples to train a robust model~\cite{Jumper2021}. In nuclear mass predictions, however, only about 2500 masses have been measured experimentally, resulting in a small and limited dataset. Moreover, nuclear mass data exhibit complex nonlinear dependencies, which further complicates the learning task. For such small-sample datasets with high complexity, conventional ML models struggle to capture the underlying complex patterns. It is therefore important to develop ML approaches that are effective for small-sample, high-complexity problems. Kolmogorov–Arnold Networks (KANs) offer a novel perspective to address this challenge. Based on the Kolmogorov‑Arnold representation theorem~\cite{russian.111.024316, Arnold2009}, KANs can represent complex multivariate functions by decomposing them into simpler univariate components, providing a theoretically grounded framework for modeling intricate systems with limited data.

    While traditional theoretical models have successfully captured the fundamental systematics of nuclear masses, a residual component of the data remains poorly described by existing approaches~\cite{STRUTINSKY1967420}. This unexplained residual—which belongs to the category of limited, high-complexity data—is crucial to address in order to improve the completeness of nuclear mass models. In this work, a simple model based on a KAN is constructed for the first time to correct the theoretical models of nuclear masses. Despite its simplicity, the model substantially improves the prediction accuracy of nuclear masses. Moreover, methods such as network visualization and feature importance analysis have been employed to enhance the model's interpretability, which further facilitates our understanding of the underlying physics.

    The paper is organized as follows. The details of KAN and data processing are discussed in Sec.~\ref{sec2}. Section~\ref{sec3} presents the application of KANs to nuclear masses. A summary and perspectives are given in Sec.~\ref{sec4}.

\section{\label{sec2}THEORETICAL FRAMEWORK}
    
\subsection{Kolmogorov-Arnold network}

Multilayer Perceptrons (MLPs) represent a foundational model in the field of artificial neural networks. Structurally, they consist of multiple layers of linear transformation modules, each followed by a nonlinear activation function. By stacking these "linear transformation $+$ nonlinear activation" layers, MLPs are capable of modeling complex nonlinear mappings. This capability is theoretically supported by the universal approximation theorem:
\begin{equation}\label{mlp}
f(\mathbf{x}) \approx \sum_{i=1}^{N(\epsilon)} a_{i} \sigma\left(\mathbf{w}_{i} \cdot \mathbf{x}+b_{i}\right)
\end{equation}
where $\mathbf{x}$ is the input vector, $N$ represents the number of neurons in the hidden layer, $\sigma$ denotes the nonlinear activation function, $a_i$ is the output weight associated with the $i$-th hidden neuron, $\mathbf{w}_i$ signifies the input weight vector of the $i$-th hidden neuron, and $b_i$ is the bias term of the $i$-th hidden neuron. Although the theorem guarantees that an MLP can approximate any continuous function given a sufficient number of neurons, its practical implementation faces limitations. The MLP's architecture, which stacks "linear transformations followed by nonlinear activations," exhibits low parameter efficiency. This results in limited efficacy in feature extraction and representation learning for high-dimensional data. Consequently, the model often requires substantial amounts of training data or more sophisticated network designs to perform learning tasks effectively. However, in scientific research, especially in fields like nuclear physics, datasets are often small but highly complex. This inherent characteristic necessitates the development of more data-efficient methods.

KANs present a novel and data-efficient approach for learning complex representations from input data. The development of KANs is inspired by the Kolmogorov-Arnold representation theorem. This theorem states that any continuous multivariate function on a bounded domain can be represented as a finite composition of continuous univariate functions and the addition operation. More specifically, for a continuous function 
\begin{equation} \label{ka}
f(\mathbf{x}) = f(x_1, \ldots, x_n) = \sum_{q=1}^{2n+1} \Phi_q \left( \sum_{p=1}^{n} \phi_{q,p}(x_p) \right)
\end{equation}
where $\phi_{q,p} \colon [0,1] \to \mathbb{R}$ and $\Phi_{q} \colon \mathbb{R} \to \mathbb{R}$. $\phi_{q,p}$ and $\Phi_{q}$ are continuous univariate functions. Equation (\ref{ka}) demonstrates that addition is the only truly multivariate operation, as all other multivariate functions can be constructed via univariate functions and summation operations. This is an encouraging finding for ML: learning a high-dimensional function is inherently equivalent to learning a number of one-dimensional functions that scales polynomially with the dimensionality. However, these one-dimensional functions may exhibit non-smooth or even fractal properties, which could pose significant challenges for practical learning algorithms. Liu \textit{et al.}~\cite{liu2025kan, 4t7t-v19l} generalized the original framework to arbitrary widths and depths. This key modification transformed the Kolmogorov-Arnold Theorem-based architecture into a trainable model.

    \begin{figure*}[htbp]
        \centering
        \includegraphics[width=0.9\textwidth]{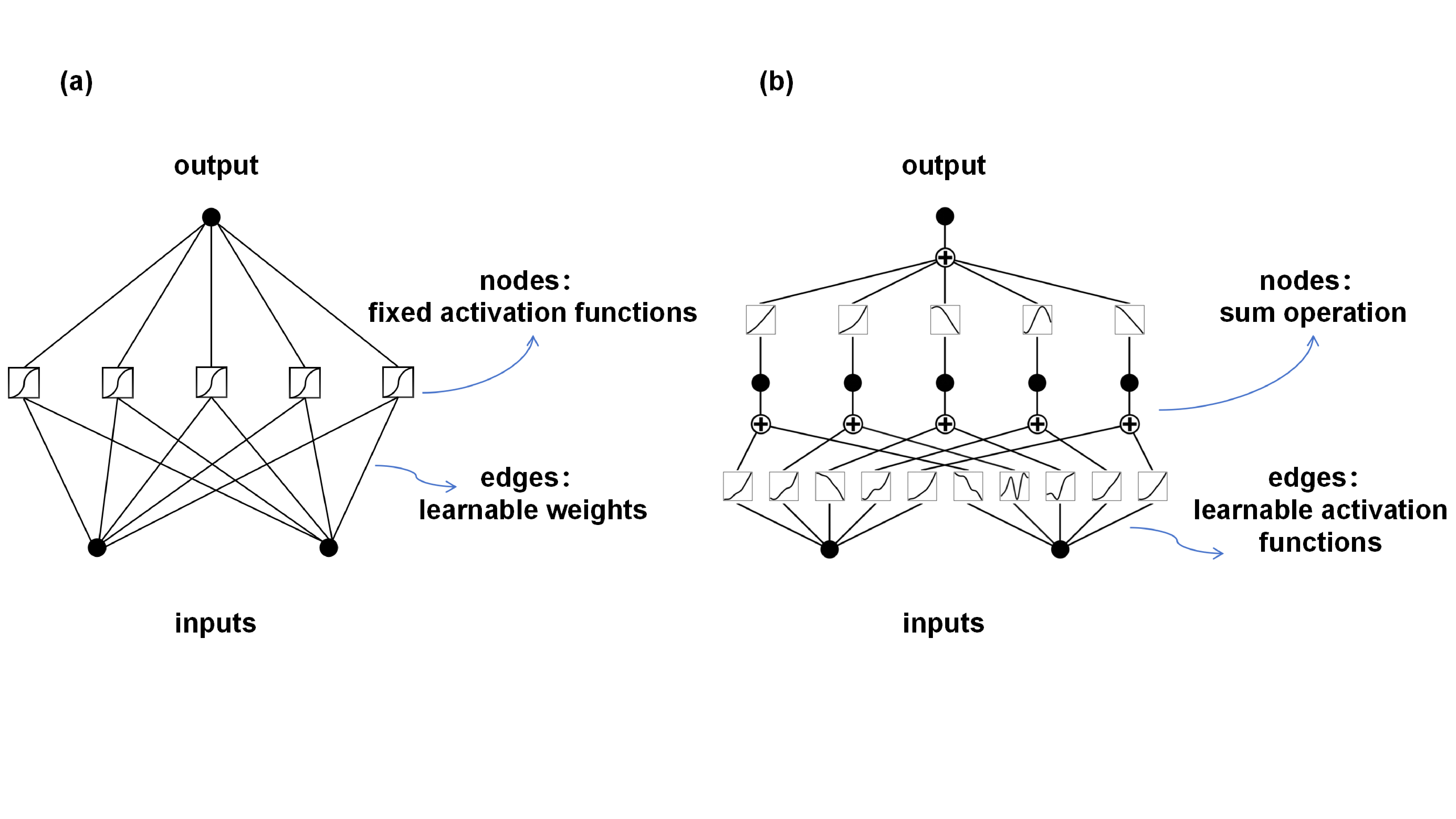}
        \caption{Comparison between (a) traditional machine learning frameworks (Multi-Layer Perceptron, MLP) and (b) Kolmogorov-Arnold Networks (KAN).}
        \label{figure1}
    \end{figure*}

Figure \ref{figure1} clearly illustrates the architectural differences between an MLP and a KAN. In the MLP, learnable weights are placed on the edges, while fixed activation functions are applied at the nodes. In contrast, the KAN employs learnable activation functions on the edges, and simple summation is performed at the nodes. Since each input undergoes a nonlinear transformation via these adaptive activation functions, the KAN exhibits a significantly enhanced capacity for feature extraction from data. The expressive power of an MLP stems primarily from architectural scaling via the accumulation of layers and nodes. Conversely, the expressive power of a KAN is inherent in the complexity of its edge functions, which can be represented by sophisticated splines. This enables even shallow KANs to achieve strong local approximation.

\subsection{Small-sample datasets with high complexity}

The nuclear mass can be decomposed into two components: a large, smooth part $M_{\text{pre}}$ that can be reliably described by nuclear models, and a residual term $M_{\delta}$ that is small but exhibits complex fluctuations, making it difficult to predict. For example, global models such as WS4~\cite{WANG2014215} accurately describe the overall trends of nuclear masses. However, they often fail to capture the residual $M_{\delta}$, which, despite its small magnitude, exhibits complex and high-frequency fluctuations.

This study aims to model the residual $M_{\delta}$—the discrepancy between theoretical predictions and experimental nuclear masses—which is essential for achieving a more accurate description. The work is based on a dataset of 2340 nuclei ($Z, N \geq 8$) selected from the AME2020 compilation. To ensure data reliability, only nuclei with experimental uncertainties not exceeding 100 keV were included. For machine learning, this represents a very small dataset. Moreover, the residuals exhibit high complexity, making this a typical case of a small-sample dataset with high complexity. Given this challenge, the KAN offers a powerful tool for modeling such data, as discussed earlier.

After careful consideration of established nuclear physics theories and the characteristics of small-sample, high-complexity data, $Z$ (proton number), $N$ (neutron number), $P$ (pairing-related quantity), and $S$ (shell-effect-related quantity) were selected as inputs for the KAN model. $Z$ and $N$ are the most fundamental variables defining a nucleus, while $P$ and $S$ are key physical quantities derived from $Z$ and $N$—corresponding to the pairing effect and shell effect in nuclear structure, respectively. Specifically, $P$ is defined as:
\begin{equation}
    P =\begin{cases}
    \frac{12}{\sqrt{A}}+\varDelta,  &\text{for\quad even-even}, \\
   -\frac{12}{\sqrt{A}}+\varDelta,  &\text{for\quad odd-odd}, \\
    \varDelta,                      &\text{for\quad odd-$A$}. \\
\end{cases}\label{eq:delta}
\end{equation}
where $A = N + Z$, and $\Delta =[(-1)^N + (-1)^Z]/2$. And $S$ is defined as:
\begin{equation}
    S=v_\mathrm{p} v_\mathrm{n}
\end{equation}
where $v_\mathrm{p}$ and $v_\mathrm{n} $ denote the differences between the proton number $Z$ and the nearest proton magic number, and between the neutron number $N$ and the nearest neutron magic number, respectively. The proton magic numbers are taken as 8, 20, 28, 50, 82 and 126, while the neutron magic numbers are 8, 20, 28, 50, 82, 126 and 184. This input design enables the KAN model to focus on learning the complex $M_{\delta}$ nonlinear patterns in the residual 
that cannot be captured by traditional theoretical models.

\section{\label{sec3}Results and discussion}

\begin{figure*}[htbp]
        \centering
        \includegraphics[width=0.9\textwidth]{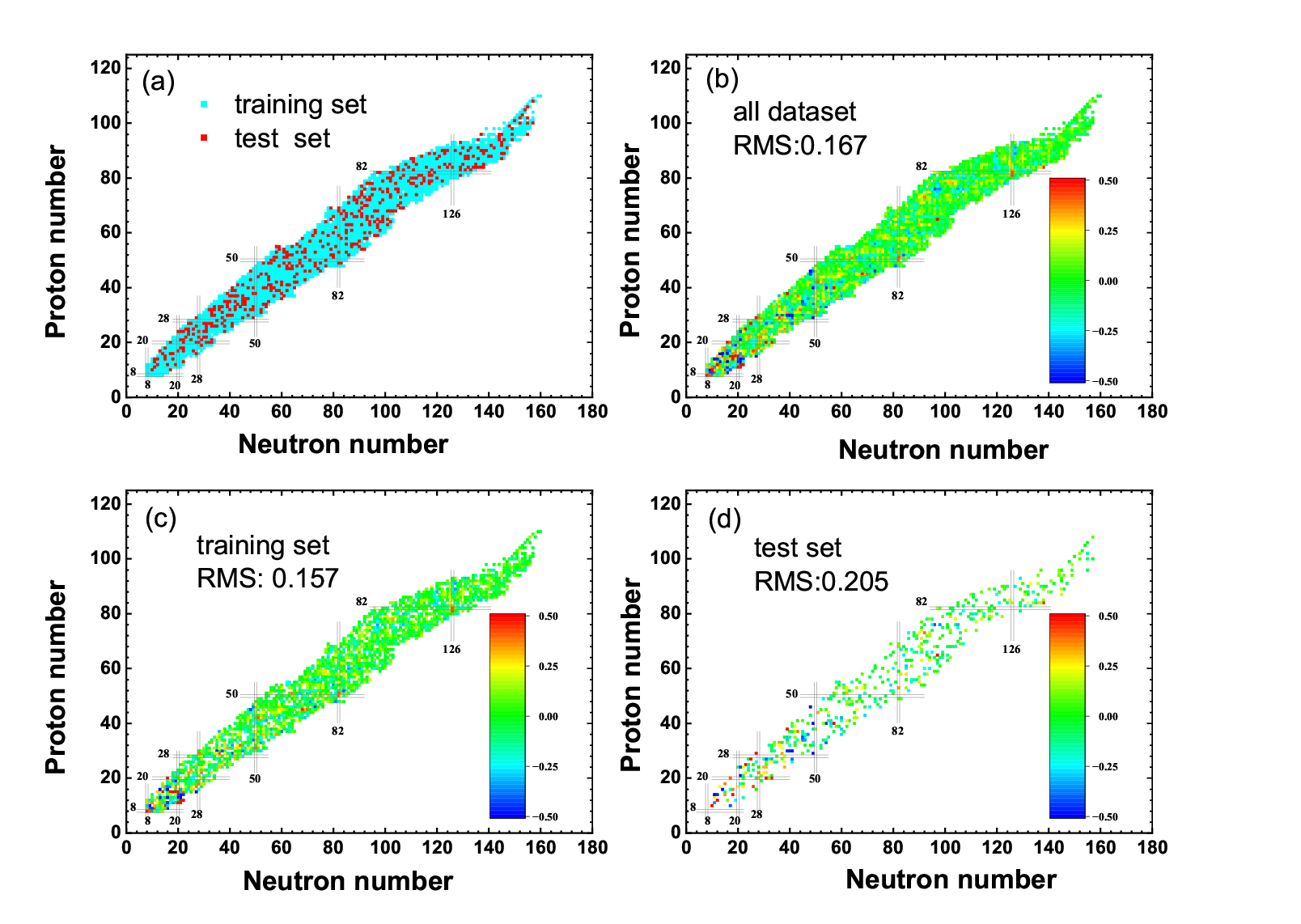}
        \caption{Differences (in MeV) between the nuclear masses predicted by KAN models and the experimental nuclear masses. (a) Data distributions of the training set and test set; (b) all dataset: (c) training set; (d) test set.}
        \label{figure2}
    \end{figure*}

Figure \ref{figure2} illustrates the deviations between the predictions of the KAN-WS4 model and experimental nuclear masses across the entire nuclide chart. To ensure a rational evaluation, the dataset was randomly split into training and test sets in an 8:2 ratio, and their distributions are shown in Fig.~\ref{figure2}(a). As presented in Fig.~\ref{figure2}(b), the root-mean-square (RMS) error over the entire nuclide chart is 0.167 MeV. Figures \ref{figure2}(c) and \ref{figure2}(d) display that the RMS values on the training set and the test set are 0.157 MeV and 0.205 MeV, respectively. The KAN-WS4 model exhibits excellent prediction accuracy across all nuclides. Compared with the classical WS4 model (RMS $\approx$ 0.3 MeV), the proposed model achieves a significant improvement in accuracy. Such high precision indicates that the model has successfully captured the main regularities of nucleon-nucleon interactions embedded in nuclear masses. It is noteworthy, however, that the predictive performance of the KAN-WS4 model is significantly inferior in the light nuclide region, while it demonstrates superior predictive capability in the medium-heavy nuclide region. This trend holds true for the entire nuclide chart, as well as the training and test sets.

\begin{figure*}[htbp]
        \centering
        \includegraphics[width=0.9\textwidth]{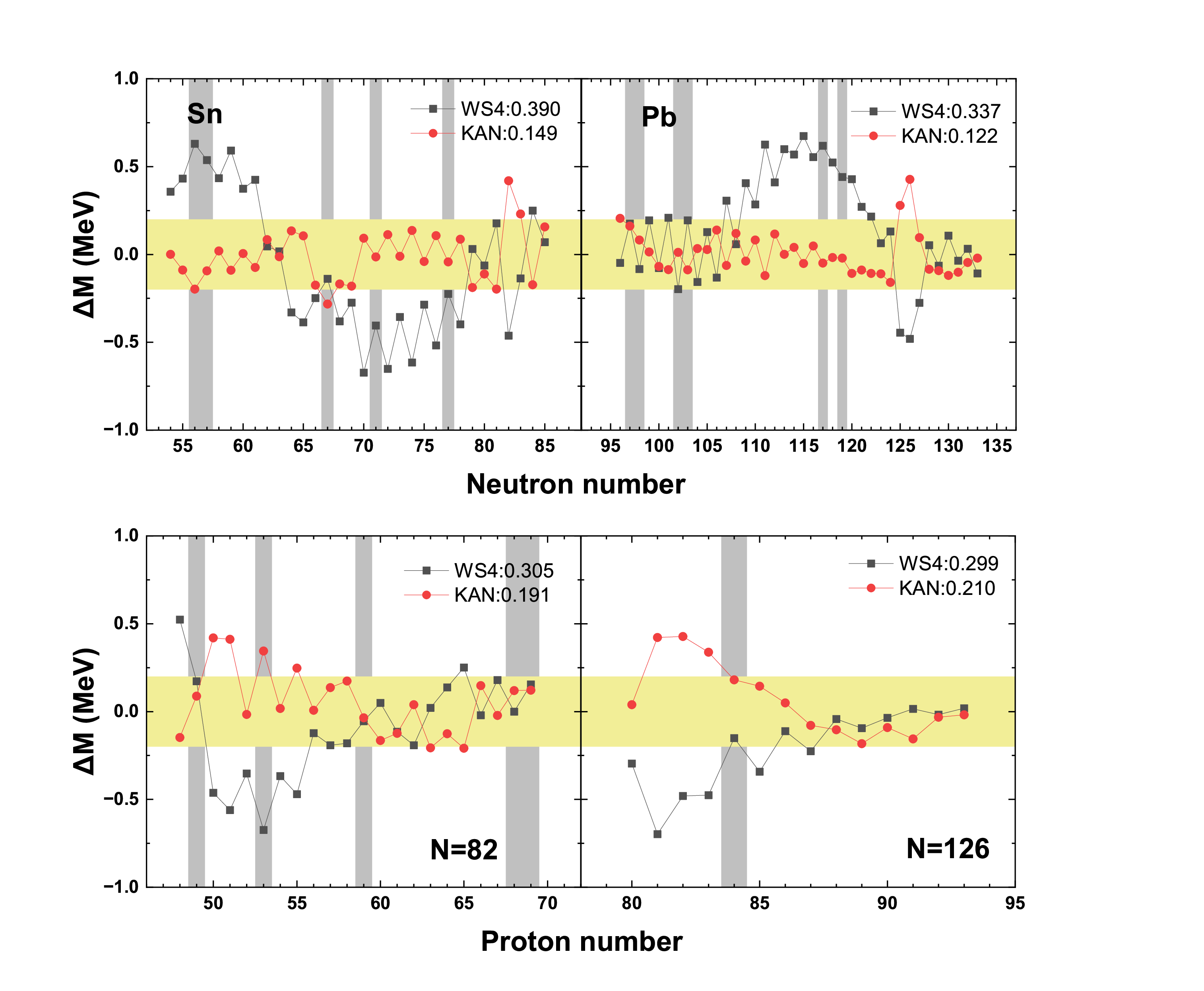}
        \caption{The training and test mass differences $\Delta M$ (in MeV) of the KAN model (red dots) with the WS4 nuclear mass model (black squares) as a reference. The shade (white) regions indicate the training (test) areas. (a) The Sn isotopic chain; (b) The Pb isotopic chain; (c) $N=82$ isotonic chain; (d) $N=126$ isotonic chain.}
        \label{figure3}
    \end{figure*}

 Figure \ref{figure2} demonstrates the high accuracy and strong generalization capability of the KAN-WS4 model from a global perspective. To reveal the model’s specific performance in key physical regions through local case analyses, four important nuclide chains associated with magic numbers are selected for individual investigation in Fig. \ref{figure3}. It should be emphasized that no deliberate selection was made of the nuclide chains with the most optimal predictions, thus endowing the results with greater reliability. Red dots represent the deviations between the KAN-WS4 model predictions and the experimental values, while black squares denote those between the WS4 model predictions and the experimental values. The white regions indicate the training set, and the black regions represent the randomly selected test set. The yellow regions correspond to errors ranging from $-$0.2 MeV to 0.2 MeV. It can be clearly observed that the prediction errors of the KAN-WS4 model are significantly smaller than those of the WS4 model across all nuclide chains. 

\begin{figure*}[htbp]
        \centering
        \includegraphics[width=0.9\textwidth]{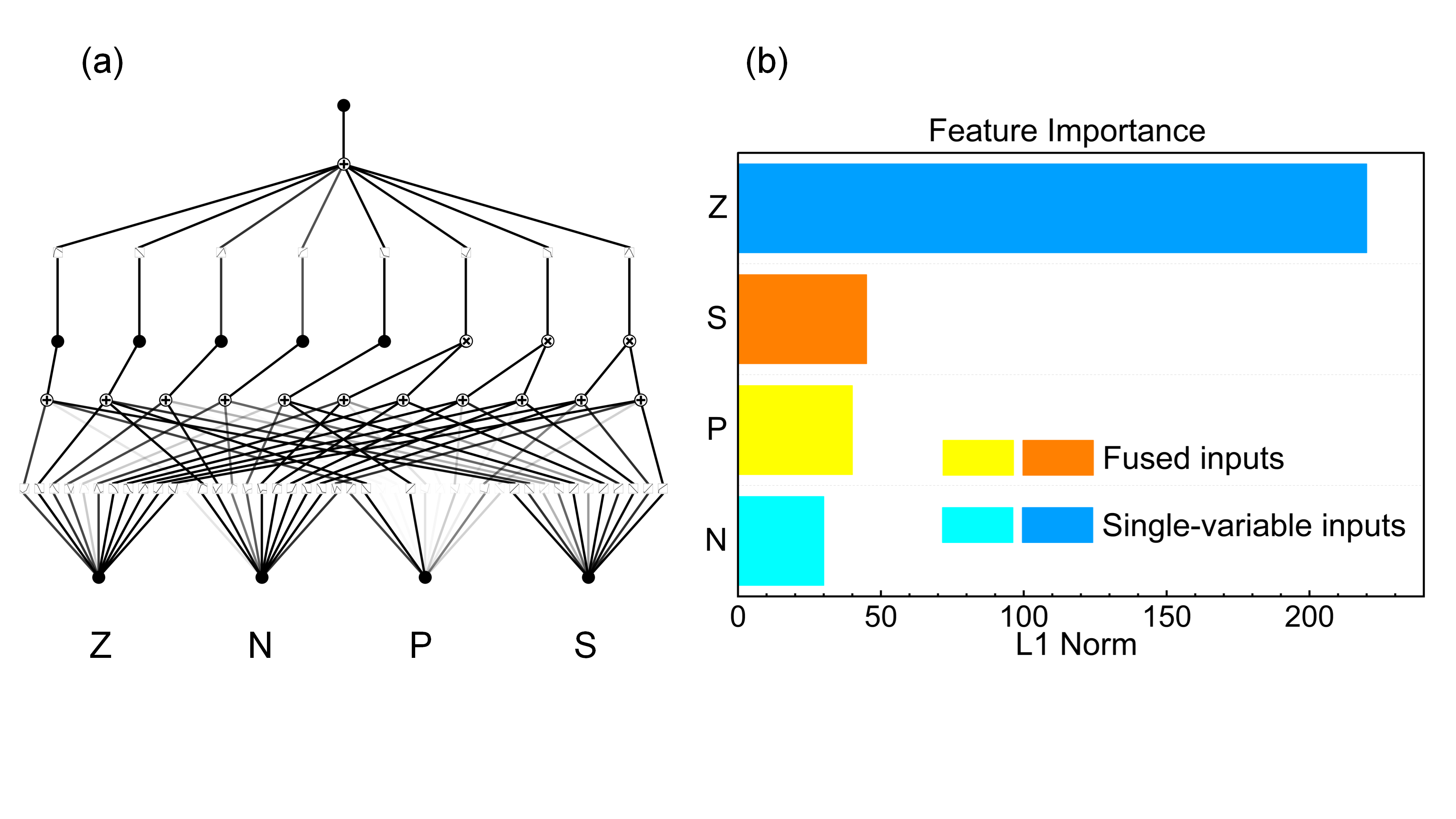}
        \caption{(a) Framework diagram of a KAN trained on the nuclear mass residual dataset, and (b) the corresponding analysis of feature importance.}
        \label{figure4}
    \end{figure*}

Figure \ref{figure4} illustrates the KAN framework employed in this study and the corresponding interpretability analysis for nuclear mass predictions. Figure \ref{figure4}(a) schematically depicts the architecture of the KAN model, which takes fundamental properties of the nucleus as inputs. The information propagates upward from four distinct inputs at the bottom layer. Through a series of learnable univariate functions that perform nonlinear transformations, the model ultimately outputs a precise prediction of the nuclear mass. The network architecture is simple, featuring just one hidden layer with few nodes, yet it delivers high predictive accuracy. This design significantly improves transparency during training and effectively addresses the common "black-box" issue in machine learning. 

Figure \ref{figure4}(b) quantifies the importance of input features by evaluating the L1 norm of the activation functions connected to each input variable. A larger L1 norm value corresponds to a greater contribution of that feature to the model output, thereby indicating a higher feature importance. Here, blue and cyan correspond to $Z$ and $N$, respectively, which are categorized as single-variable inputs—these represent independent inputs of individual physical quantities. In contrast, orange and yellow correspond to $S$ and $P$, respectively, which are fused inputs—these are composite inputs derived from the integration of $Z$ and $N$. The results indicate that the proton number ($Z$) is the most significant feature in determining nuclear mass residuals. The shell-effect-related quantity ($S$) and the pairing-related quantity ($P$) rank as the second and third most important features, respectively, surpassing even the neutron number ($N$) in influence. 

The identification of $Z$ as the most critical feature by the KAN-WS4 model likely reflects its role in capturing and correcting systematic deviations in the Coulomb energy term of the WS4 model. This finding further underscores that the proton number serves as a nexus of multiple nucleonic interactions, defining nuclear structure and acting as a primary source of systematic errors in existing global theoretical models. The importance of $S$ and $P$ closely following that of $Z$ indicates that, after accounting for the global framework determined by $Z$, microscopic corrections related to shell and pairing effects constitute the next most significant factors. The relatively lower importance of $N$ does not imply it is physically insignificant; rather, it suggests that its systematic contributions may already be well captured by the theoretical model, or that its information is partially embedded in other features such as $S$ and $P$. As a result, the independent information provided solely by $N$ in the remaining residuals is comparatively limited.

Given the significant influence of $Z$, the operational mechanism of the KAN-WS4 model essentially involves learning the complex nonlinear relationship between the proton number and mass residuals, thereby diagnosing and correcting deficiencies in existing theoretical models—particularly in $Z$-sensitive terms such as the Coulomb energy, proton shell effects, and symmetry energy. This data-driven ranking of feature importance significantly improves the transparency and credibility of the KAN model. More importantly, it delivers a novel perspective for understanding the origins of nuclear mass residuals through data-driven analysis. The results indicate that the key to improving future nuclear mass models likely lies in a more thorough consideration of the proton number $Z$ and a more precise characterization of the microscopic mechanisms related to shell structure and nucleon pairing correlations.

To further demonstrate the model's stability, the KAN framework was applied to systematically optimize and correct five other distinct nuclear mass theoretical models, namely Bhagwat~\cite{PhysRevC.90.064306}, DZ28~\cite{PhysRevC.52.R23}, FRDM~\cite{PhysRevLett.108.052501}, HFB31~\cite{PhysRevC.93.034337}, and KTUY~\cite{10.1143/PTP.113.305}, as summarized in Table \ref{table1}. The results demonstrate that the prediction accuracy of all these models was significantly improved after the KAN framework learned the corresponding mass residuals. Among these models, the WS4 and Bhagwat models achieved remarkably high accuracy after KAN-based correction. This indicates that these models themselves have already well described the dominant components of nuclear masses, and the unaccounted physical effects contained in their residuals are relatively amenable to being learned by the KAN framework. Notably, some models with large initial errors, such as KTUY and FRDM, exhibited a substantial reduction in errors after KAN-based correction, demonstrating the robust capability of this method in rectifying the systematic biases of the original models. Taken together, as a highly interpretable machine learning architecture, the KAN is widely applicable to enhancing the accuracy of various types of nuclear mass models, demonstrating significant application potential in theoretical nuclear physics research.

\begin{table}
\centering  
\renewcommand{\arraystretch}{1.3}
\footnotesize
\caption{Six theoretical models optimized via KAN. $\sigma_{\text{pre}}$ denotes the RMS error between the predictions of the bare theoretical mass model and the experimental data, whereas $\sigma_{\text{KAN}}$ represents the RMS error between the predictions of the theoretical mass model refined by the KAN and the experimental data. All quantities are given in MeV.}\label{table1}
\begin{tabular}{c c c c}  
    \hline 
    \hline 
    \makebox[0.12\textwidth][c]{Model} &  
    \makebox[0.13\textwidth][c]{$\sigma_{\text{pre}}$} & 
    \makebox[0.1\textwidth][c]{$\sigma_{\text{KAN}}$(train)} & 
    \makebox[0.12\textwidth][c]{$\sigma_{\text{KAN}}$(test)} \\
    \hline 
    WS4 & 0.286 & 0.157 & 0.205 \\
    Bhagwat & 0.303  & 0.164  & 0.214  \\
    DZ28 & 0.407 & 0.174 & 0.227 \\
    FRDM & 0.586 & 0.208 & 0.273 \\
    HFB31 & 0.567 & 0.340 & 0.356 \\
    KTUY & 0.715 & 0.173 & 0.234 \\
    \hline
    \hline
\end{tabular}
\end{table}

\section{\label{sec4}SUMMARY AND PROSPECTS}

This study introduces the KAN, a novel machine learning architecture, into nuclear mass prediction, presenting a powerful tool for addressing the challenge of small-sample datasets with high complexity. Based on the theoretical framework that decomposes nuclear mass into a smooth global component (described by models like WS4) and complex fluctuating residuals, a simple KAN model was constructed specifically to learn the residuals between theoretical predictions and experimental values, forming an efficient KAN-WS4 model. This model enhanced the precision of the WS4 model from approximately 0.3 MeV to about 0.16 MeV, demonstrating high predictive accuracy. In contrast to the "black-box" nature of traditional neural networks, KAN's structure is simple, transparent, and offers strong interpretability. Through feature importance analysis, the study identified the proton number as the most critical feature determining the mass residuals. This data-driven discovery suggests potential systematic biases in existing global theoretical models (e.g., WS4) related to proton-number-dependent terms, such as coulomb energy and proton shell effects, providing valuable insights for future model refinement. Furthermore, applying the KAN framework to five other mass models, yielded significant accuracy improvements, demonstrating its broad applicability and potential for correcting various nuclear mass models.

While KAN's simplicity enables symbolic regression for deriving interpretable expressions, its current utility is largely restricted to rediscovering simple physical laws~\cite{liu2025kan}. True discovery of laws in complex physical systems via ML remains a distant goal, particularly for small-sample, high-complexity datasets where minor deviations are easily obscured by error propagation. This work presents a new approach to studying such datasets. In future work, the complex residual patterns learned by KAN could be fed back and integrated into physical theoretical models themselves. For example, KAN-derived insights may refine or construct more accurate phenomenological formulas (e.g., for coulomb energy, symmetry energy), advancing the use of AI to discover physical laws. The KAN framework could also be applied to other key nuclear physics problems—such as predicting nuclear charge radii, decay lifetimes, and fission barriers—all of which face small-sample, high-complexity challenges. KAN’s interpretability advantage is expected to yield new physical insights in these areas.

\begin{acknowledgments}
This work is supported by the National Natural Science Foundation of China (Grant No. 12475119), the Key Laboratory of Nuclear Data Foundation (JCKY2025201C154), the JSPS Grant-in-Aid for Scientific Research (S) under Grant No.~20H05648,
the RIKEN iTHEMS Program, and the RIKEN Pioneering Project: Evolution of Matter in the Universe.
\end{acknowledgments}

\nocite{*}

\bibliography{apssamp}

\end{document}